\documentclass[onecolumn]{elsart3p}
\bibstyle{elsart-num}

\usepackage{graphicx}

\usepackage{amssymb}

\newcommand{\li}{{\rm Li}}
\newcommand{\R}{{\rm R}}

\begin{document}

\begin{frontmatter}



\title{Is there quantum chaos in the prime numbers?}


\author{Todd K. Timberlake},
\author{Jeffery M. Tucker}

\address{Department of Physics, Astronomy, \& Geology, Berry College, Mount Berry, GA  30149-5004 USA}

\begin{abstract}
A statistical analysis of the prime numbers indicates possible traces of quantum chaos.  We have computed the nearest neighbor spacing distribution, number variance, skewness, and excess for sequences of the first $N$ primes for various values of $N$.  All four statistical measures clearly show a transition from random matrix statistics at small $N$ toward Poisson statistics at large $N$.  In addition, the number variance saturates at large lengths as is common for eigenvalue sequences.  This data can be given a physical interpretation if the primes are thought of as eigenvalues of a quantum system whose classical dynamics is chaotic at low energy but regular at high energy.  We discuss some difficulties with this interpretation in an attempt to clarify what kind of physical system might have the primes as its quantum eigenvalues.
\end{abstract}

\end{frontmatter}

\section{Introduction}
\label{intro}

One of the primary results that has emerged from the study of the quantum dynamics of classically chaotic systems is the connection between eigenvalue statistics and classical dynamics \cite{Stockmann1999,Haake2001,Reichl2004}.  Quantum systems with classically regular dynamics typically have eigenvalues that follow Poisson statistics \cite{Berry1977}.  Quantum systems with classically chaotic dynamics have eigenvalues that follow the statistics of random matrix eigenvalues \cite{Bohigas1984,Bohigas1985}.  Different ensembles of random matrices correspond to different symmetries of the system.  The eigenvalues of time-reversible systems without spin-1/2 interactions follow the statistics of the Gaussian Orthogonal Ensemble (GOE), the eigenvalues of time-irreversible systems follow the statistics of the Gaussian Unitary Ensemble (GUE), and the eigenvalues of systems with spin-1/2 interactions follow the Gaussian Symplectic Ensemble (GSE).  Generic quantum systems, which have mixed chaotic and regular classical dynamics, follow statistics that are intermediate between Poisson and random matrix statistics \cite{Berry1984}.

This surprising connection between quantum eigenvalues and random matrices has led to an even more surprising connection between quantum mechanics and number theory.  The imaginary parts of the non-trivial zeros of the Riemann zeta function, a function of paramount importance in the theory of prime numbers, display GUE statistics to very high precision \cite{Odlyzko1987}.  This finding led to speculation that the zeta zeros can be thought of as eigenvalues of a time-irreversible quantum system that is classically chaotic.

We take a more direct approach to connecting prime numbers and quantum physics by applying to the primes themselves the same statistical analyses that are used to study eigenvalue sequences.  Some attempts in this direction have already been published.  Porter examined the nearest neighbor spacing distribution (NNSD), the distribution of differences between consecutive primes, for primes in the vicinity of $1.2 \times 10^8$ \cite{Porter1965}.  The resulting NNSD generally follows the expectation for Poisson statistics, but with two major deviations: large peaks at spacings equal to multiples of 6, and a shortage of spacings near zero.  The first deviation can be explained by the fact that all primes greater than 3 are equal to either $1 (\bmod \ 6)$ or $-1 (\bmod \ 6)$.  The second deviation can be partially attributed to the fact that, with one exception, all primes are odd integers and so the spacing between consecutive primes cannot be less than 2.  Liboff and Wong also studied the NNSD of primes, but they examined primes in the range 2 to $10^7$ \cite{Liboff1998}.  They came to quite a different conclusion from that of Porter, claiming that the primes seem to fit the GOE distribution rather than the Poisson. There are some problems with their analysis in that they did not unfold the primes (see Section \ref{levelspacings}) before calculating the NNSD and the fit of their histograms to the GOE distribution is questionable.  Their results are, however, sufficient to call into question the conclusion that the primes (at least the small primes) follow Poisson statistics.  

In this paper we present a statistical study of both relatively large ($\approx 3\times 10^{13}$) and small primes.  Several statistical studies of prime numbers have been published recently \cite{Wolf1996,Wolf1997,Szpiro2004,Ares2006,Szpiro2007} but these studies did not use the tools that are typically applied to quantum eigenvalue spectra.  Using these tools can provide information on what kind of physical system might have the primes as its quantum eigenvalues because of the connection between eigenvalue statistics and classical dynamics.  Our goals are to shed some light on this intriguing topic, clarify the conflicting results of Porter and Liboff and Wong, and add to the growing body of statistical data on the primes.

\section{Level spacings}
\label{levelspacings}

Once a particular subsequence of primes has been generated it must be unfolded before the statistical calculations are performed.  If the primes are thought of as energy eigenvalues then unfolding accounts for the way the density of states changes with energy.  The standard procedure for unfolding a sequence of eigenvalues $x_i$ is to insert each eigenvalue into the average level staircase function, $\bar{\eta}(x)$.  The average level staircase function is a smooth function that gives the approximate number of eigenvalues less than $E$.  For the primes the exact level staircase function is usually denoted by $\pi(x)$ and the average level staircase function is given by a smooth function that approximates $\pi(x)$.  The celebrated Prime Number Theorem indicates that $\pi(x)$ is well-approximated by $x/\log(x)$ in the limit $x \to \infty$.  However, a better approximation for $\pi(x)$ is given by the log integral function:
\begin{equation}
\label{li}
\li(x) = \int_{2}^{x} \frac{dt}{\log(t)}.
\end{equation}
For small primes an even better approximation for $\pi(x)$ is given by 
\begin{equation}
\label{rfunc}
\R(x) = \sum_{m=0}^{\lfloor \log(x)/\log(2) \rfloor} \frac{\mu(m)}{m} \li\left(x^{1/m}\right),
\end{equation}
with the Moebius function $\mu(m)$ defined by
\begin{equation}
\label{mu}
\mu(m) = \left\{
	\begin{array}{ll}
	1, & \mbox{if $m=1$}\\
	0, & \mbox{if $m$ is divisible by a square of a prime}\\
	(-1)^k, & \mbox{otherwise}
	\end{array} \right.
\end{equation}
where $k$ is the number of prime divisors of $m$.  To unfold the primes we transform from the sequence of primes $p_n$ to the sequence $\epsilon_n = R(p_n)$.  The sequence $\epsilon_n$ will then have a mean density of one throughout the sequence.  Making the mean density uniform in this way allows us to more easily study fluctuations about the mean.

\begin{figure}[h]
\includegraphics[bb = 126 225 477 576, width=5in, height=5in]{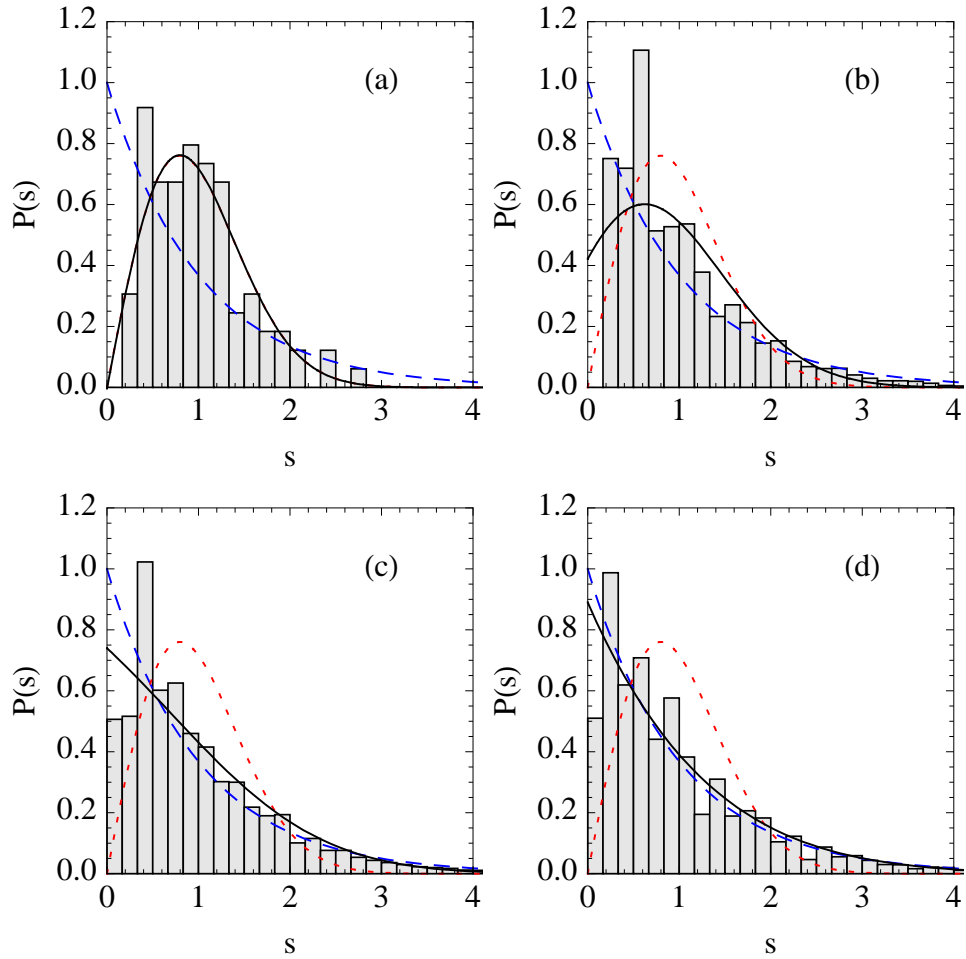}
\caption{\label{spacings} Nearest neighbor spacing distributions (NNSD) for subsequences of unfolded primes.  The first three histograms show the NNSD for the first (a) $10^2$, (b) $10^4$, and (c) $10^6$ primes.  The fourth histogram (d) shows the NNSD for the first $10^6$ primes after the $10^{12}$th prime.  The curves show the Poisson (dashed) and GOE (dotted) distributions as well as the Berry-Robnik distribution (solid) obtained by fitting the number variance data (see Fig. \ref{numvar}).}
\end{figure}

One of the most common ways to study these fluctuations is to examine the NNSD mentioned above.  Figure \ref{spacings} shows the NNSD for four different subsequences of primes, along with the curves for the Poisson distribution and the Wigner GOE distribution (which closely approximates the exact GOE distribution) given by
\begin{eqnarray}
P_{\mathrm{Poisson}}(s) = e^{-s} & \mbox{and} & P_{\mathrm{GOE}}(s) = \frac{\pi s}{2} e^{-\pi s^2/4}.
\end{eqnarray}
Note that the most likely spacing for Poisson statistics is $s=0$, while the most likely spacing for GOE statistics is $s=\sqrt{2/\pi}$.  For this reason eigenvalues that follow Poisson statistics are said to exhibit \emph{level clustering} while those that follow GOE (or any of the random matrix) statistics are said to exhibit \emph{level repulsion}.  Figure \ref{spacings}a shows the NNSD for the first 100 primes, which shows definite level repulsion and closely resembles GOE statistics.  Increasing the number of primes in the subsequence, as in Figs. \ref{spacings}b and \ref{spacings}c, reduces the level repulsion and reveals a transition toward Poisson statistics.  Fig. \ref{spacings}d shows the NNSD for the  $10^6$ primes immediately following the $10^{12}$th prime.  The histogram in Fig. \ref{spacings}d fits the Poisson distribution except for deviations similar to those observed by Porter.  Overall these results suggest that small primes exhibit level repulsion and obey GOE statistics, while larger primes show a progressive trend toward level clustering and Poisson statistics.

Berry and Robnik developed a statistical model that interpolates between GOE and Poisson statistics \cite{Berry1984}.  They considered an eigenvalue sequence that is a combination of one or more independent Poisson sequences and a single independent GOE sequence.  The spacing distribution for such a sequence is given by
\begin{equation}
\label{BRspacings}
P(s,\rho_1) = \rho_1^2 e^{-\rho_1 s} \mathrm{erfc}(\sqrt{\pi}\bar{\rho}s/2) + (2\rho_1\bar{\rho} + \pi\bar{\rho}^3s/4) \exp(-\rho_1s - \pi\bar{\rho}^2s^2/4)
\end{equation}
where $\mathrm{erfc}(x)$ is the complementary error function, $\rho_1$ is the fraction of the eigenvalues that come from any of the Poisson sequences, and $\bar{\rho} = 1-\rho_1$ is the fraction of the eigenvalues that come from the GOE sequence.  The solid curves in Fig. \ref{spacings} show the Berry-Robnik distribution with the parameter $\rho_1$ determined by a fit to the number variance data (see Section \ref{other}) for each sequence.  The curves fit the histograms reasonably well except for deviations similar to the deviations from Poisson statistics seen in Porter's work mentioned above.  Note that the Berry-Robnik curve in Fig. \ref{spacings}a is indistinguishable from the GOE curve and the Berry-Robnik curve in Fig. \ref{spacings}d is very close to the Poisson curve.

\section{Other Statistics}
\label{other}

The NNSD provides information about correlations between neighboring values in a sequence.  To more closely examine the statistical properties of primes we can turn to other statistical measures that provide information about longer-range correlations, or correlations between more than just two values in the sequence.  Three commonly used statistical measures are the number variance, skewness, and excess (or kurtosis).  The number variance ($\Sigma^2$) measures two-point correlations over a longer range than the NNSD while the skewness ($\gamma_1$) and excess ($\gamma_2$) measure 3- and 4-point correlations, respectively.  Each of these statistics can be defined in terms of the moments
\begin{equation}
\label{moments}
\mu_j = \left\langle \left(n - \langle n \rangle\right)^j \right\rangle
\end{equation}
where $n$ counts the number of values in an interval of length $L$ and $\langle \ldots \rangle$ represents an average taken over many such intervals throughout the entire sequence.  In terms of these moments
\begin{equation}
\label{statseqn}
\begin{array}{lll}
\Sigma^2 = \mu_2, & \gamma_1 = \mu_3 \mu_2^{-3/2}, & \gamma_2 = \mu_4 \mu_2^{-2}-3.
\end{array}
\end{equation}
The expected results for these statistics in the Poisson, GOE, and GUE cases can be found in Chapter 16 of Ref. \cite{Mehta2004}.

\begin{figure}[ht]
\includegraphics[bb = 126 225 477 576, width=5in, height=5in]{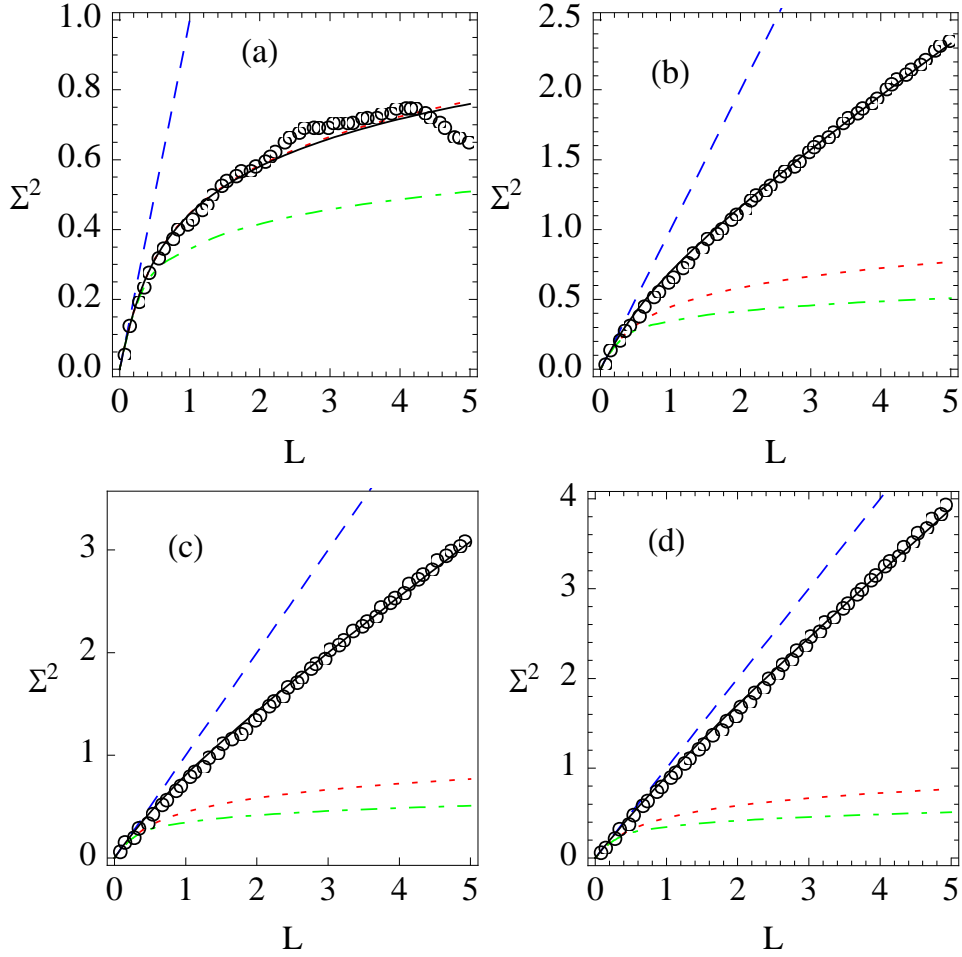}
\caption{\label{numvar} Number variance $\Sigma^2$ as a function of interval length $L$ for subsequences of unfolded primes.  The open circles show the data for the first (a) $10^2$, (b) $10^4$, and (c)$10^6$ primes as well as for (d) the first $10^6$ primes after the $10^{12}$th prime.  The curves show the expected results for Poisson (dashed), GOE (dotted), and GUE (dot-dashed) statistics as well as the Berry-Robnik distribution (solid) that best fits the data.  Note the varying scales on the $\Sigma^2$-axis.}
\end{figure}

Figure \ref{numvar} shows the number variance for the same subsequences of primes examined in Fig. \ref{spacings}, as well as the curves for Poisson, GOE, and GUE statistics.  The data for the first 100 primes  in Fig. \ref{numvar}a matches the GOE curve (with small deviations at larger values of $L$ that are likely due to poor statistics since there are so few numbers in this sequence).  The data for the first $10^4$ (Fig. \ref{numvar}b) and $10^6$ (Fig. \ref{numvar}c) primes show a shift away from the GOE curve and toward the Poisson curve as more primes are included in the subsequence.  The data for the first $10^6$ primes after the $10^{12}$th prime (Fig. \ref{numvar}d) continues this trend, but still displays significant deviations from the Poisson curve.  

The solid curves in Fig. \ref{numvar} are derived by fitting the data to the Berry-Robnik distribution mentioned in Section \ref{levelspacings}.  The formula for the number variance in Berry-Robnik statistics is
\begin{equation}
\label{BRnv}
\Sigma^2_{\mathrm{BR}} (L, \rho_1) = \Sigma^2_{\mathrm{Poisson}}(\rho_1 L) + \Sigma^2_{\mathrm{GOE}}(\bar{\rho} L)
\end{equation}
where again it is assumed that a fraction $\rho_1$ of the eigenvalues in the sequence follow Poisson statistics while the remaining fraction ($\bar{\rho} = 1-\rho_1$) follow GOE statistics.  It is clear from Fig. \ref{numvar} that the Berry-Robnik distribution fits the data quite well.  Table \ref{rho1} shows the values of $\rho_1$ determined by fitting the number variance data in the range $0 < L \leq 5$ to the Berry-Robnik distribution for several different subsequences of primes.  These data clearly show a transition from GOE statistics for small primes toward Poisson statistics for larger primes.  However, it should be noted that $\rho_1$ is still well below one even for primes near the $10^{12}$th prime.

\begin{figure}[ht]
\includegraphics[bb = 153 243 495 567, width=5in, height=4.7in]{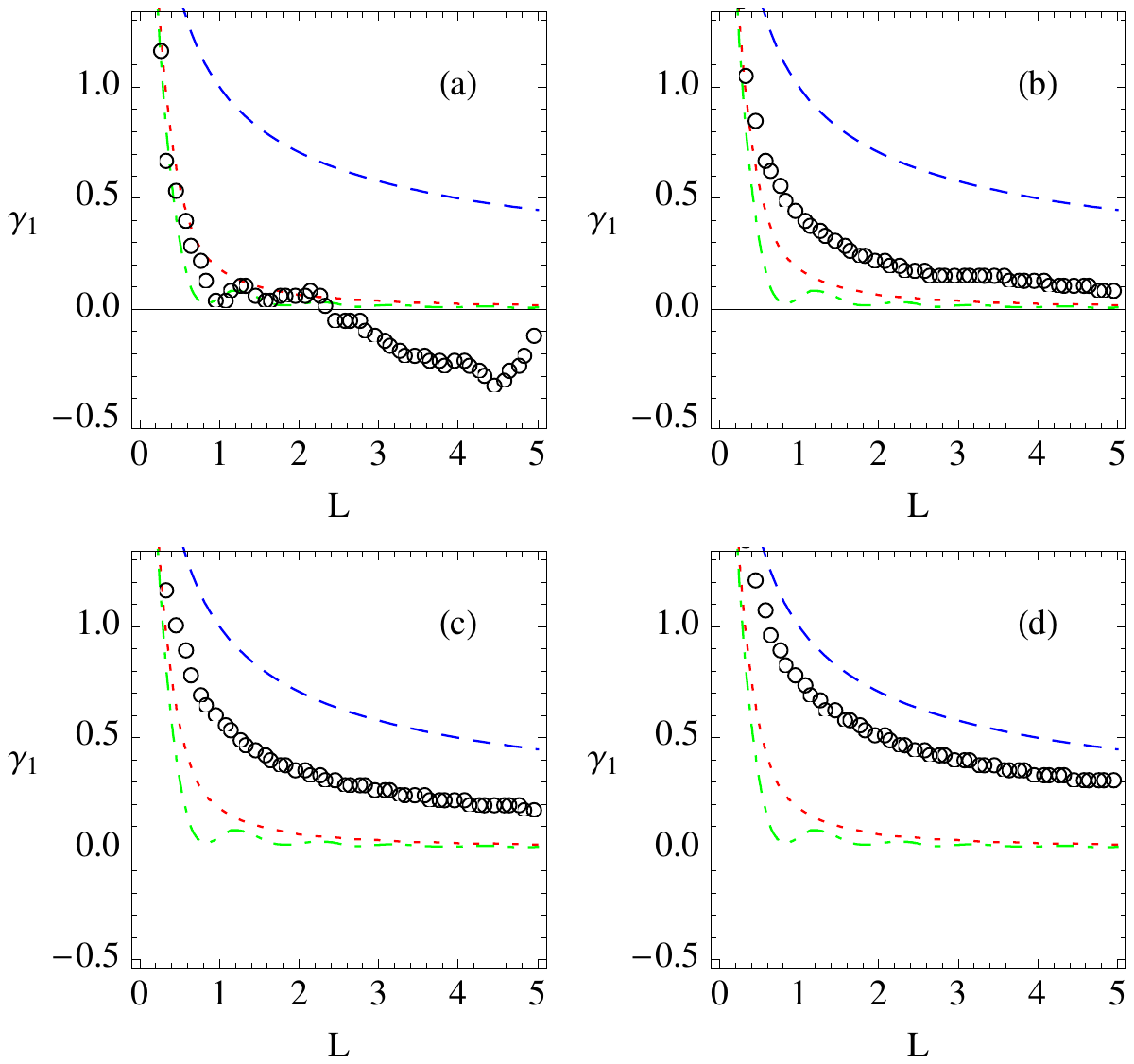}
\caption{\label{skew} Skewness $\gamma_1$ as a function of interval length $L$ for subsequences of unfolded primes.  The open circles show the data for the first (a) $10^2$, (b) $10^4$, and (c)$10^6$ primes as well as for (d) the first $10^6$ primes after the $10^{12}$th prime.  The curves show the expected results for Poisson (dashed), GOE (dotted), and GUE (dot-dashed) statistics.}
\end{figure}

\begin{figure}[ht]
\includegraphics[bb = 153 243 495 567, width=5in, height=4.7in]{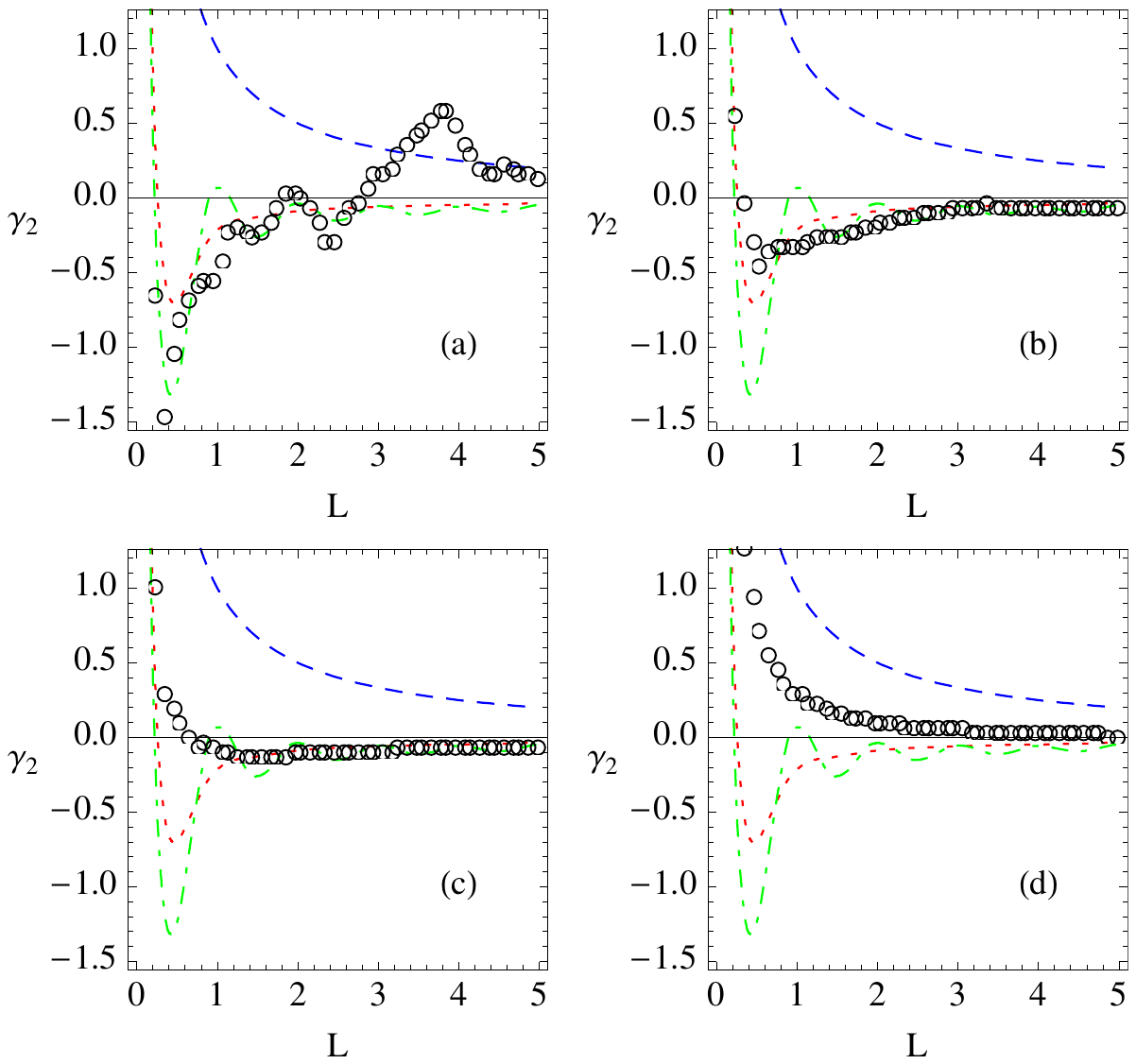}
\caption{\label{exc} Excess (kurtosis) $\gamma_2$ as a function of interval length $L$ for subsequences of unfolded primes.  The open circles show the data for the first (a) $10^2$, (b) $10^4$, and (c)$10^6$ primes as well as for (d) the first $10^6$ primes after the $10^{12}$th prime.  The curves show the expected results for Poisson (dashed), GOE (dotted), and GUE (dot-dashed) statistics.}
\end{figure}

Higher-order correlations follow a pattern similar to that seen in the number variance.   Figure \ref{skew} shows the skewness as a function of $L$ for the same sequences examined in Figures \ref{spacings} and \ref{numvar}.   Figure \ref{exc} shows the excess for these sequences.  The curves in both of these figures show the expected results for Poisson, GOE, and GUE statistics.  In both figures there is a transition from random matrix statistics for small primes toward Poisson statistics for larger primes.  However, the skewness for the first 100 primes (Fig. \ref{skew}a) seems to fall somewhere between the GOE and GUE curves and the excess for the first 100 primes (Fig. \ref{exc}a) appears to fit the GUE curve better than the GOE curve.  It is difficult to determine the significance of the skewness and excess results for the first 100 primes because this data shows large fluctuations due to the small number of values in the sequence.  The skewness and excess curves for the Berry-Robnik distribution using the $\rho_1$ values from Table \ref{rho1} do not fit the data in Figs. \ref{skew} and \ref{exc} and we were unable to obtain an acceptable fit to these data using any value for $\rho_1$.  This could be an indication that the primes are not really a combination of independent Poisson and GOE sequences, or it could indicate that more than one GOE sequence is involved.  We attempted to fit the skewness and excess data to a Berry-Robnik distribution for a mixture of Poisson and GUE sequences and for a mixture of Poisson and two GOE sequences without any success.

\begin{figure}[ht]
\includegraphics[bb = 126 225 477 576, width=5in, height=5in]{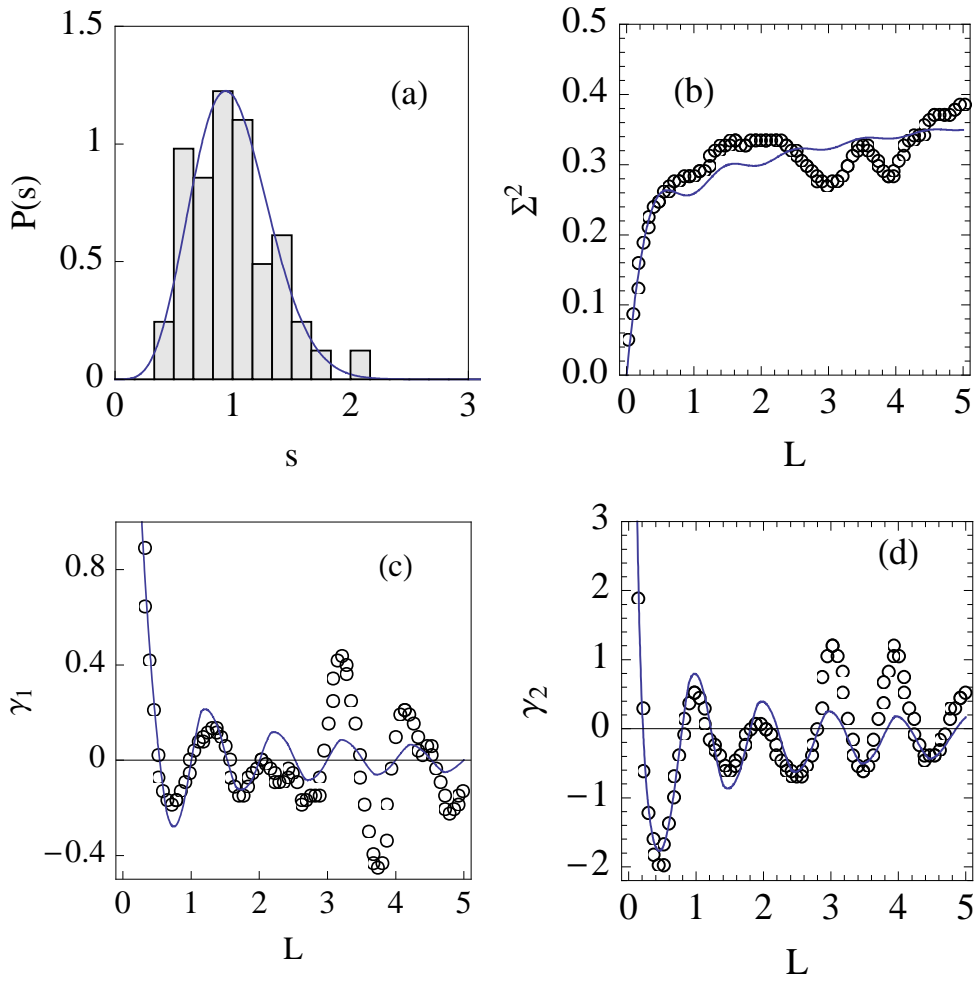}
\caption{\label{alternate} Statistical results for the first 50 alternate primes ($2, 5, 11, \ldots, 523$).  The histogram in (a) shows the NNSD for this sequence, while the open circles in (b)-(d) show the number variance, skewness, and excess, respectively.  The solid curve in all four plots represents the expected result for GSE statistics.}
\end{figure}

To further clarify whether or not the first 100 primes really follow GOE statistics we also examined the statistics of the first 50 alternate primes ($2, 5, 11, \ldots, 523$).  If every other number in a GOE sequence is removed, the resulting sequence is known to follow GSE statistics.  We generated a list of the first 50 alternate primes, unfolded this sequence using $\R(x)$, and rescaled the resulting sequence so that the mean spacing was one.  We then examined the level spacing distribution, number variance, skewness, and excess as described above.  The results are presented in Figure \ref{alternate}.    The solid curves in Figure \ref{alternate} show the expected results for the GSE distribution.  The Wigner GSE level spacing distribution is given by 
\begin{equation}
\label{GSEspacings}
P_{\mathrm{GSE}}(s) = \frac{2^{18} s^4}{3^6 \pi^3} e^{-64s^2/(9\pi)}
\end{equation}
and the expected results for $\Sigma^2$, $\gamma_1$, and $\gamma_2$ for GSE statistics can be found in Chapter 16 of Ref. \cite{Mehta2004}.  The fit between the data and the GSE results is fairly good in all cases, although there are noticeable differences that may simply be fluctuations that result from the small number of values in the sequence.  Overall, the statistics of the first 50 alternate primes adds support to the claim that the first 100 primes follow GOE statistics.

\section{Interpretation of the results}
\label{interpret}

These statistical results can be given a physical interpretation because of the connection between classical dynamics and eigenvalue statistics discussed in Section \ref{intro}.  Numerical studies of several model systems have shown similar transitions between random matrix and Poisson statistics \cite{Wintgen1987,Lin1989}.  In these cases the transition in statistics occurs when a parameter is changed in such a way as to change the classical dynamics of the system from chaotic to regular.  In the case of the primes the transition occurs as one moves toward larger and larger primes.  If the primes are thought of as energy eigenvalues of a quantum system, then the classical dynamics of that system would seem to be chaotic at low energies and increasingly regular at higher energies.  There are, however, some problems with this interpretation which we will discuss below.

One potential problem is that the deviation from Poisson statistics might be due to improper unfolding.  The use of some local unfolding procedures on Poisson sequences has been shown to produce apparent deviations from Poisson statistics \cite{Gomez2002}.  However, the unfolding procedure we have used is a global unfolding based on the function $R(x)$ which is known to accurately approximate the mean level staircase function for the primes.  We have also tried unfolding the sequence of primes using $x/\log(x)$ and $\li(x)$ and there is no substantial change in the statistical results, so we are confident that our unfolding procedure is correct.

\begin{figure}[ht]
\includegraphics[bb = 0 0 165 105, width=5in, height=3.2in]{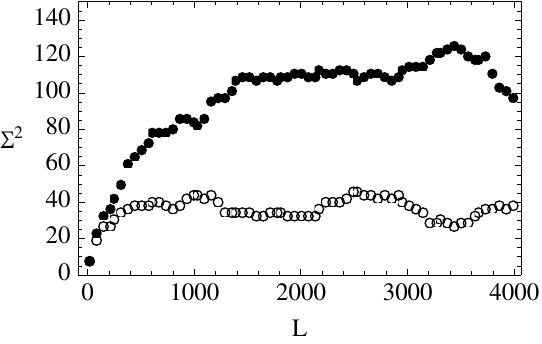}
\caption{\label{sat} Number variance $\Sigma^2$ as a function of interval length $L$ for subsequences of unfolded primes.  The points show the data for primes numbered 1 to 10,000 (open circles) and for primes numbered 10,001 to 20,000 (filled circles).}
\end{figure}

Another alternative to our suggested interpretation is that the primes may be eigenvalues of an integrable system and the deviations from Poisson statistics at low energies may be due to the influence of short periodic orbits in the corresponding classical system.  Berry has shown, using semiclassical arguments, that the spectral rigidity (a statistic closely related to the number variance) will saturate at a value of $L$ determined by the period of the shortest periodic orbit of the classical system \cite{Berry1985}.  This effect can lead to large deviations from Poisson statistics in integrable systems, even mimicking GOE statistics, at low energies while at higher energies the statistics approach the Poisson curves \cite{Casati1985,Drozdz1991}.  The usual pattern in these cases is that the spectral rigidity (or the number variance) follows the Poisson curve up to $L \approx L_{\rm max}$, then saturates and remains at a roughly fixed value as $L$ is increased beyond $L_{\rm max}$.  Since $L_{\rm max}$ increases with energy the overall result is a transition toward Poisson statistics at higher energies.  In the primes, however, we see deviations at small $L$ even when the energies are quite large (as shown in Fig. \ref{numvar}d).  In fact, the number variance for the primes does saturate but at values of $L$ much larger than those shown in Fig. \ref{numvar}.  Figure \ref{sat} shows that this saturation of the number variance occurs for sequences of primes and that the value of $L$ at which the number variance saturates increases as the size of the primes in the sequence increases.

There are other issues that complicate the association between eigenvalue statistics and classical dynamics.  Large deviations from Poisson statistics have been found in pseudointegrable systems \cite{Lewenkopf1990,Cheon1991}.  In addition, fully chaotic systems can display deviations from random matrix statistics due to the presence of localized quantum states associated with classical structures like cantori \cite{Wintgen1988} or a continuous line of periodic orbits \cite{Shudo1990}.

One fact that seems to contradict the suggested interpretation of the prime statistics is that if the primes are eigenvalues of a quantum system then that system must be one-dimensional, as pointed out by Mussardo \cite{Mussardo1997}.  This conclusion follows from the fact that the primes grow roughly like $n \log(n)$.  Eigenvalues of simple quantum wells with potentials of the form $V(x) = |x|^{k}$ in $d$ dimensions grow like $n^{2k/(d(k+2))}$ according to Weyl's law \cite{Haake2001}.  This means the steepest possible growth is for a billiard system with hard walls (corresponding to $k \to \infty$), which gives $n^{2/d}$.  So eigenvalues of systems with two (or more) spatial dimensions cannot grow faster than $n$.  This result seems to rule out the possibility that primes are eigenvalues of a pseudointegrable or chaotic system since all time-independent 1D systems are integrable.   However, simple one-dimensional quantum wells do not follow Poisson statistics but rather display the non-generic statistics characteristic of harmonic oscillators.  It seems we are left with no options that fit the statistical data on the primes.

We would like to suggest two alternatives that merit further research.  The first is that the primes might be eigenvalues of a one-dimensional potential well that is more complicated than the simple wells discussed above.  One such potential has already been suggested \cite{Mussardo1997}.  If this is the case then the apparent random matrix statistics for small primes has no relation to chaotic classical dynamics.   We are unaware, though, of any one-dimensional system with eigenvalues that exhibit random matrix statistics at low energies and Poisson statistics at high energies.  Another possibility to consider is that the primes could be eigenvalues of a one-dimensional potential well subject to a periodic driving field.  Periodically-driven anharmonic wells can exhibit exactly the type of classical motion suggested by our statistical analysis, namely chaos at low energies and regular motion at higher energies (see Ref. \cite{Lin1986} for an example).  However, for time-periodic systems the appropriate eigenvalues are quasienergies rather than energies.  Quasienergies are only defined modulo $\hbar\omega$, where $\omega=2\pi/T$ and $T$ is the period of the time-dependent part of the system's Hamiltonian.  Thus, numerical calculations produce quasienergies that are all within a single Brillouin zone between $0$ and $\hbar\omega$.  It is possible to assign each quasienergy to its proper Brillouin zone by tracing the evolution of each eigenvalue as the strength of the time-periodic perturbation is increased from zero, but it is not at all clear how this would affect the statistical properties of the sequence.

\section{Conclusions}

We have analyzed subsequences of prime numbers using statistical tools used in the study of quantum eigenvalues.  Our results clearly show level repulsion among the small primes with a progressive tendency toward Poisson statistics for larger primes.  The statistical data suggest that primes could be eigenvalues of a quantum system whose classical counterpart is chaotic at low energies but increasingly regular at higher energies.  There are, however, some difficulties with such an interpretation  particularly because the primes would have to be eigenvalues of a one-dimensional system.

In spite of the difficulties involved in their interpretation, we feel that these results strongly suggest a connection between primes and eigenvalues of a quantum system.  These statistical results may also be of interest from a purely mathematical perspective.  The deviation from Poisson statistics for small primes might be related to other ``atypical'' behaviors exhibited by small primes.  For example, for sequences of the first $n$ primes the number of primes equal to $1 (\bmod \ 6)$ is always less than or equal to the number of primes equal to $-1 (\bmod \ 6)$ up to $n \approx 6 \times 10^{11}$ (a phenomenon known as Chebyshev's bias) even though the infinite sequence contains equal amounts of both types of primes.  A more dramatic example is that the $\li(x) > \pi(x)$ for $x < 10^{316}$, even through Littlewood proved that these two functions must cross each other an infinite number of times.  Our results seem to indicate a gradual convergence toward Poisson statistics as more primes are included in the sequence.  It may be that the infinite sequence of (unfolded) primes follows Poisson statistics, even though primes near the $10^{12}$th prime show significant deviations from Poisson statistics.  However, we are unaware of any rigorous proof that the full sequence of unfolded primes follows Poisson statistics.







\newpage

\section{Figures}

\newpage

\newpage

\newpage

\newpage

\newpage

\newpage

\section{Tables}

\begin{table}[h]
\begin{center}
\begin{tabular}{|c|c|c|c|}\hline
First $n$ primes & $\rho_1$ & First $10^6$ primes after $k$ & $\rho_1$ \\ \hline
$n=10^2$ & -0.00181 & $k=10^7$ & 0.555921 \\ \hline
$n=10^3$ & 0.239504 & $k=10^8$ & 0.585383 \\ \hline
$n=10^4$ & 0.328879 & $k=10^9$ & 0.61471 \\ \hline
$n=10^5$ & 0.430437 & $k=10^{10}$ & 0.633034 \\ \hline
$n=10^6$ & 0.489928 & $k=10^{11}$ & 0.652538 \\ \hline
 & & $k=10^{12}$ & 0.668721 \\ \hline
 \end{tabular}
 \end{center}
 \caption{\label{rho1} Values of $\rho_1$, the fit parameter for the Berry-Robnik distribution, obtained by fitting the number variance data for various sequences of primes.  The left two columns show the results for sequences of primes beginning with the first primes and having various lengths.  The right two columns show the results for sequences of one million primes beginning at different starting values.}
 \end{table}


\begin{thebibliography}{10}
\expandafter\ifx\csname url\endcsname\relax
  \def\url#1{\texttt{#1}}\fi
\expandafter\ifx\csname urlprefix\endcsname\relax\def\urlprefix{URL }\fi

\bibitem{Stockmann1999}
H.-J. St\"{o}ckmann, Quantum Chaos: An Introduction, Cambridge University
  Press, New York, 1999.

\bibitem{Haake2001}
F.~Haake, Quantum Signatures of Chaos, 2nd Edition, Springer, New York, 2001.

\bibitem{Reichl2004}
L.~E. Reichl, The Transition to Chaos: Conservative Classical Systems and
  Quantum Manifestations, 2nd Edition, Springer, New York, 2004.

\bibitem{Berry1977}
M.~V. Berry, M.~Tabor, Proc. Roy.
  Soc. Lond. A 356 (1977) 375.

\bibitem{Bohigas1984}
O.~Bohigas, M.~J. Giannoni, C.~Schmit, Phys. Rev. Lett. 52
  (1984) 1.

\bibitem{Bohigas1985}
O.~Bohigas, R.~U. Haq, A.~Pandey, Phys. Rev. Lett. 54 (1985) 1645.

\bibitem{Berry1984}
M.~V. Berry, M.~Robnik, J. Phys. A 17 (1984) 2413.

\bibitem{Odlyzko1987}
A.~M. Odlyzko, Math. Comp. 48 (1987) 273.

\bibitem{Porter1965}
C.~E. Porter, in: C.~E. Porter (Ed.),
  Statistical Theories of Spectra: Fluctuations, Academic Press, 1965, p.
  2.

\bibitem{Liboff1998}
R.~L. Liboff, M.~Wong, Int. J. Th. Phys. 37 (1998) 3109.

\bibitem{Wolf1996}
M.~Wolf, in:
  A.~M. P.~Borcherds, M.~Bubak (Ed.), Proceedings of the Eighth Joint EPS-APS
  International Conference, Krakow, 1996, p. 361.

\bibitem{Wolf1997}
M.~Wolf, Physica A 241 (1997)
  493.

\bibitem{Szpiro2004}
G.~G. Szpiro, Physica A 341 (2004) 607.

\bibitem{Ares2006}
M.~C. S.~Ares, Physica A 360 (2006) 285.

\bibitem{Szpiro2007}
G.~G. Szpiro, Physica A 384 (2007) 291.

\bibitem{Mehta2004}
M.~L. Mehta, Random Matrices, 3rd Edition, Elsevier, San Diego, 2004.

\bibitem{Wintgen1987}
D.~Wintgen, H.~Friedrich, Phys. Rev. A 35
  (1987) 1464.

\bibitem{Lin1989}
W.~A. Lin, L.~E. Reichl, Phys. Rev. A 40 (1989) 1055.

\bibitem{Gomez2002}
J.~M.~G. G\'{o}mez, R.~A. Molina, A.~R. {n}o, J.~Retamosa, Phys. Rev. E 66 (2002) 036209.

\bibitem{Berry1985}
M.~V. Berry, Proc. Roy. Soc. Lond. A
  400 (1985) 229.

\bibitem{Casati1985}
G.~Casati, B.~V. Chirikov, I.~Guarneri, Phys. Rev. Lett. 54 (1985) 1350.

\bibitem{Drozdz1991}
S.~Dro\.{z}d\.{z}, J.~Speth, Phys. Rev. Lett. 67 (1991) 529.

\bibitem{Lewenkopf1990}
C.~H. Lewenkopf, Phys. Rev. A 42 (1990) 2431.

\bibitem{Cheon1991}
T.~Cheon, T.~Mizusaki, T.~Shigehara, N.~Yoshinaga, Phys. Rev. A 44 (1991) R809.

\bibitem{Wintgen1988}
D.~Wintgen, H.~Marxer, Phys.
  Rev. Lett. 60 (1988) 971.

\bibitem{Shudo1990}
A.~Shudo, Y.~Shimizu, Phys. Rev. A 42 (1990) 6264.

\bibitem{Mussardo1997}
G.~Mussardo, cond-mat/9712010 (1997).

\bibitem{Lin1986}
W.~A. Lin, L.~E. Reichl, Physica D 19 (1986) 145.

\end{thebibliography}
\end{document}